\documentclass[prd,amsmath,amssymb,longbibliography,superscriptaddress,twocolumn,nofootinbib,10pt]{revtex4-1}

\pdfoutput=1
\usepackage{graphicx}
\usepackage{dcolumn}
\usepackage{bm}
\usepackage{amssymb}
\usepackage{latexsym}
\usepackage{booktabs}
\usepackage{amsmath}
\usepackage{multirow}
\usepackage{array}

\usepackage[colorlinks=true, linkcolor=blue, citecolor=blue]{hyperref}

\begin{document}

\title{The open-Universe signal: A model artifact rather than genuine curvature}

\author{Peng-Ju Wu} \email{wupengju@nxu.edu.cn}
\affiliation{School of Physics, Ningxia University, Yinchuan 750021, China}

\begin{abstract}
Recent late-Universe observations suggest an open Universe. If confirmed, such a departure from spatial flatness would carry profound implications for our understanding of cosmic inflation and the ultimate fate of the Universe. Motivated by this intriguing result and the release of new data, we revisit the question using baryon acoustic oscillation measurements from DESI DR2, multiple Type Ia supernova samples, refined strong gravitational lensing time-delay analyses, and the most up-to-date cosmic chronometer data. We find that within the $\Lambda$ cold dark matter ($\Lambda$CDM) paradigm, the combined data still prefer an open Universe with $\Omega_K=0.049\pm0.037$. However, this preference vanishes in extensions to $\Lambda$CDM, where the data instead favor a flat Universe. The model comparison shows that for $\Lambda$CDM, introducing new physics is preferred over merely allowing spatial curvature, and flat $\Lambda$CDM extensions perform better than their curved counterparts. We therefore argue that the mild open-Universe signal is an artifact of limited model flexibility, rather than a genuine feature of late-Universe observations.
\end{abstract}

\maketitle
\section{Introduction}
Whether the Universe is open, flat, or closed is among the most fundamental questions in cosmology. A non-zero spatial curvature, parameterized by the curvature parameter $\Omega_K$, would directly challenge cosmic inflation (which predicts near-perfect flatness) and alter expectations regarding the long-term evolution and ultimate fate of the Universe. Precisely constraining the sign and magnitude of $\Omega_K$ is therefore essential for refining our comprehensive understanding of the Universe.

This question has gained renewed attention from high-precision measurements of the cosmic microwave background (CMB) \cite{Aghanim:2018eyx,Park:2017xbl,Handley:2019tkm,DiValentino:2019qzk,Efstathiou:2020wem}. Analyses of Planck CMB temperature and polarization anisotropies favor a closed Universe, yielding $\Omega_K=-0.044_{-0.015}^{+0.018}$ \cite{Aghanim:2018eyx}. Including CMB lensing information gives $\Omega_K=-0.0106\pm0.0065$, which still favors a closed Universe but with reduced significance \cite{Aghanim:2018eyx}. More recent Planck PR4 likelihoods further alleviate this preference, yielding $\Omega_K=-0.012\pm0.010$ \cite{Rosenberg:2022sdy,Tristram:2023haj}. Thus far, the tendency toward $\Omega_K<0$ has not disappeared. These results are obtained within the $\Lambda$ cold dark matter ($\Lambda$CDM) framework. If this deviation cannot be attributed to residual systematics, it may signal new physics, such as modified early dynamics that leave a genuine curvature imprint, or dynamical dark energy whose effects are absorbed into the curvature term.

Constraints on $\Omega_K$ from CMB suffer from a geometric degeneracy, which can be broken by baryon acoustic oscillation (BAO) observations \cite{Zaldarriaga:1997ch,Efstathiou:1998xx,Bond:1997wr,Vagnozzi:2020rcz,Vagnozzi:2020dfn,Dhawan:2021mel,Efstathiou:2020wem}. Combining Planck with BAO data yields $\Omega_{K}=0.0007\pm0.0019$, consistent with spatial flatness \cite{Aghanim:2018eyx}. However, concerns have been raised regarding the robustness of such joint analysis, with some studies suggesting the existence of a ``curvature tension'' between early- and late-Universe datasets \cite{DiValentino:2019qzk}. Combining newer DESI BAO data with CMB yields $\Omega_{K}=0.0024\pm0.0016$, hinting at an open Universe \cite{DESI:2024mwx}. This suggests a pattern: early-Universe data tend toward $\Omega_K<0$, while late-Universe data favor $\Omega_K>0$. Their combination thus gives a near-zero value---not necessarily because both datasets confirm flatness, but perhaps because their respective biases partially cancel each other out. Supporting this, Wu and Zhang \cite{Wu:2024faw} used only late-Universe probes and found $\Omega_K = 0.106 \pm 0.056$ within the $\Lambda$CDM model, indicating a preference for an open Universe. Similar conclusions appear in Refs.~\cite{Jiang:2024xnu,Song:2025ddm,Yadav:2025wbc,Chaudhary:2025bfs}. Of course, some studies report that late-Universe observations are consistent with flatness \cite{Comini:2026nsj,Du:2025csv,Liu:2024yib,Fortunato:2025qxc}.

In this paper, we leverage the latest late-Universe observations to test whether the preference for an open Universe is a genuine physical signal or merely an artifact of the minimal $\Lambda$CDM framework. Currently, $\Omega_K$ constraints are most commonly derived within $\Lambda$CDM. However, growing evidence supports new physics beyond $\Lambda$CDM \cite{Gialamas:2025pwv,Shajib:2025tpd,Giare:2024smz,Silva:2025hxw,Shah:2025ayl,Pan:2025qwy,Li:2024qso,Urena-Lopez:2025rad,Escudero:2024uea,Du:2024pai,Du:2025xes,Chen:2025wwn,Wang:2025zri,Li:2025dwz,Yang:2025mws,Odintsov:2024woi,Chudaykin:2024gol,Yang:2024kdo,Wu:2025vrl}, and $\Omega_K$ is strongly degenerate with the new physics parameters. Thus, while $\Lambda$CDM-based constraints provide a baseline, they lack persuasiveness due to the model's restrictive assumptions. To address this limitation and obtain more comprehensive $\Omega_K$ constraints, we consider four distinct, physically motivated extensions of $\Lambda$CDM: (i) dark matter with a relaxed equation of state (EoS), allowing non-zero dark matter pressure (denoted $\Lambda$nDM); (ii) dynamical dark energy with a free EoS ($w$CDM), allowing departure from the cosmological constant; (iii) non-gravitational interaction between dark energy and dark matter (IDE), parameterized by a coupling term \cite{Billyard:2000bh,Amendola:1999qq}; (iv) holographic dark energy (HDE), motivated by the holographic principle \cite{Cohen:1998zx,Li:2004rb}. We constrain $\Omega_K$ within each framework and quantify how parameter degeneracies affect the results, thereby strengthening the robustness of conclusions.

\section{Method and Data}
We now consider the non-flat counterparts of the models introduced above, adding the prefix ``o'' (o$\Lambda$CDM, o$\Lambda$nDM, o$w$CDM, oIDE, oHDE). Their Hubble parameters are defined as follows.

(\romannumeral1) \textbf{o$\Lambda$CDM:}
\begin{align}\label{oLCDM}
H(z)=H_0\sqrt{\Omega_{\rm m}(1+z)^3+\Omega_K(1+z)^2+\Omega_{\rm de}},
\end{align}
where $H_0$ is the Hubble constant, and $\Omega_{\rm m}$ and $\Omega_{\rm de}=1-\Omega_{\rm m}-\Omega_K$ are the density parameters of matter (including baryons and cold dark matter) and dark energy, respectively.

(\romannumeral2) \textbf{o$\Lambda$nDM:}
\begin{align}\label{oLnDM}
H(z) = &H_0 \Big[ \Omega_{\text{dm}}(1+z)^{3(1+w_{\rm dm})} +\Omega_{\rm b}(1+z)^3   \nonumber \\
& +\Omega_K(1+z)^2 + (1 - \Omega_{\text{dm}}- \Omega_{\text{b}}-\Omega_K) \Big]^{1/2},
\end{align}
where $\Omega_{\rm dm}=\Omega_{\rm m}-\Omega_{\rm b}$ is the dark matter density parameter and $w_{\rm dm}$ is the dark matter EoS parameter.

(\romannumeral3) \textbf{o$w$CDM:}
\begin{align}\label{owCDM}
H(z) = &H_0 \Big[ \Omega_{\text{m}}(1+z)^3 +\Omega_K(1+z)^2 \nonumber \\
& + (1 - \Omega_{\text{m}}-\Omega_K)(1+z)^{3(1+w_{\rm de})} \Big]^{1/2},
\end{align}
where $w_{\rm de}$ is the dark energy EoS parameter.

(\romannumeral4) \textbf{oIDE:} If there exists an interaction between dark energy and dark matter, the energy conservation equations can be written as
\begin{align}\label{IDEQ}
\dot{\rho}_{\rm de}+3H(1+w_{\rm de}){\rho}_{\rm de}=+Q, \nonumber \\
\dot{\rho}_{\rm dm}+3H(1+w_{\rm dm}){\rho}_{\rm dm}=-Q,
\end{align}
where ${\rho}_{\rm de}$ and ${\rho}_{\rm dm}$ are the energy densities of dark energy and dark matter, respectively, the dot denotes the derivative with respect to time, and $Q$ is the energy transfer rate, for which we adopt $Q=\beta H \rho_{\rm de}$, where $\beta$ is a dimensionless coupling parameter; $\beta > 0$ indicates dark matter decays into dark energy, $\beta < 0$ indicates dark energy decays into dark matter, and $\beta = 0$ corresponds to no interaction. For cold dark matter and the cosmological constant, the conservation equations yield
\begin{align}
&H(z) =H_0\left[\Omega_{\rm m}(1 + z)^3+ \Omega_K(1+z)^2 + \right. \nonumber \\
& \left.(1 - \Omega_{\rm m}-\Omega_K) \left( \frac{\beta}{\beta+3} (1 + z)^3 + \frac{3}{\beta+3} (1 + z)^{-\beta} \right)\right]^{1/2}.
\end{align}

(\romannumeral5) \textbf{oHDE:} For dark energy density defined with the future event horizon~\cite{Li:2004rb}, the evolution equations are
\begin{align}
&\frac{1}{H(z)} \frac{{\rm d}H(z)}{{\rm d}z} =-\frac{\Omega_{\rm de}(z)}{1+z}  \nonumber \\
& \times \left[\frac{1}{2} + \frac{\sqrt{\Omega_{\rm de}(z) + c^2 \Omega_K(z)}}{c}+ \frac{\Omega_K(z)-3}{2\,\Omega_{\rm de}(z)}\right], \nonumber \\
& \frac{{\rm d}\Omega_{\rm de}(z)}{{\rm d}z}=-\frac{2\,\Omega_{\rm de}(z)\,[1 - \Omega_{\rm de}(z)]}{1+z} \nonumber \\
& \times \left[\frac{1}{2} + \frac{\sqrt{\Omega_{\rm de}(z) + c^2 \Omega_K(z)}}{c}- \frac{\Omega_K(z)}{2\,[1 - \Omega_{\rm de}(z)]}\right],
\end{align}
where $\Omega_{K}(z)$ and $\Omega_{\text{de}}(z)$ denote the corresponding density parameters at redshift $z$.

We constrain cosmological models using the latest data from four probes: DESI DR2 BAO measurements \cite{DESI:2025zgx} (treating the sound horizon at drag epoch as a free parameter to avoid any dependence on early-Universe observations); three type Ia supernova (SN) samples ( DES-Dovekie \cite{DES:2025sig}, Pantheon+ \cite{Brout:2022vxf}, and Union3 \cite{Rubin:2023jdq}) with the absolute magnitude marginalized over; cosmic chronometer (CC) $H(z)$ measurements (32 data points from \cite{Moresco:2022phi} supplemented by three additional points from DESI \cite{Loubser:2025fzl}); and eight strong gravitational lensing time delay (TD) systems from the TDCOSMO collaboration \cite{TDCOSMO:2025dmr,Birrer:2020tax,Hogg:2023khs} with a full hierarchical likelihood analysis that consistently accounts for lens mass and line-of-sight uncertainties.

\section{Results and Discussions}
We constrain cosmological parameters using the Markov chain Monte Carlo method with uniform priors: $H_0 \in [20,\,100] \, \mathrm{km/s/Mpc}$, $\Omega_{\rm m} \in [0.01,\,0.99]$, $\Omega_{\rm b} \in [0.01,\,0.2]$, $\Omega_K \in [-1,\,1]$, $r_{\rm d} \in [100,\,200] \, \mathrm{Mpc}$, $w_{\rm dm} \in [-0.3,\,0.3]$, $w_{\rm de} \in [-3,\,1]$, $\beta \in [-2,\,2]$, and $c \in [0.5,\,1.5]$. The dark matter density parameter $\Omega_{\rm dm}$ is not sampled directly but derived as $\Omega_{\rm dm}=\Omega_{\rm m} - \Omega_{\rm b}$. No CMB priors are imposed on $r_{\rm d}$ or $\Omega_{\rm b}$, ensuring that our results are independent of early-Universe observations. Convergence is ensured by the Gelman-Rubin statistic with $R-1<0.01$ for all parameters. The $1\sigma$ errors for the marginalized parameter constraints are summarized in Table~\ref{tab:results}.

\begin{figure*}
\includegraphics[scale=0.6]{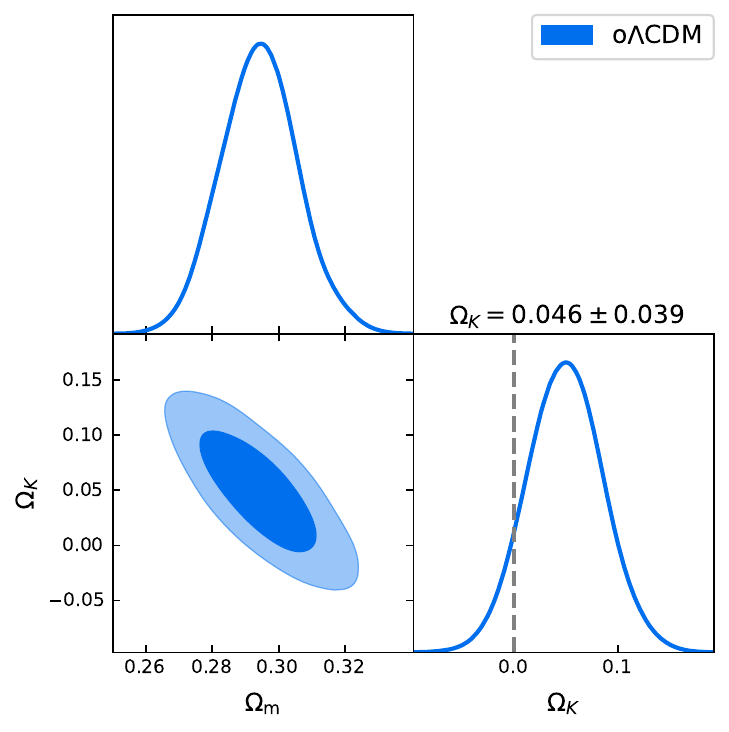}
\includegraphics[scale=0.49]{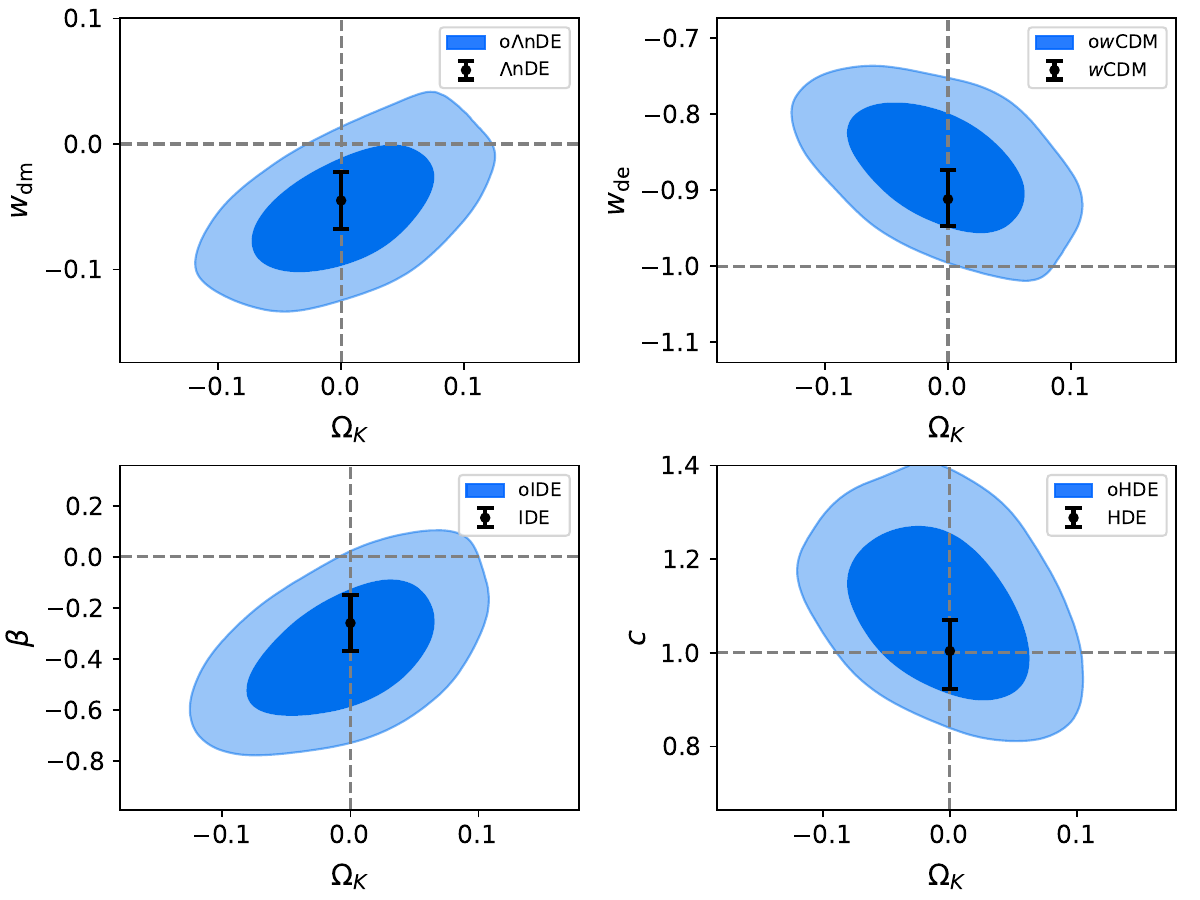}
\caption{Constraints on $\Omega_K$ in o$\Lambda$CDM and its extensions using the BAO+SN+CC+TD data. The grey vertical dashed line corresponds to the flat Universe, and the horizontal lines represent the following reference values: cold dark matter with $w_{\rm dm}=0$ (for the o$\Lambda$nDM model), the cosmological constant with $w_{\rm de}=-1$ (for the o$w$CDM model), no interaction between dark energy and dark matter with $\beta=0$ (for the oIDE model), and the phantom divide for dark energy with $c=1$ (for the oHDE model), respectively.}
\label{Contours}
\end{figure*}

\begin{table*}[!htb]
\caption{Cosmological constraints from the BAO+SN+CC+TD data. We consider the flat $\Lambda$CDM model and its extensions including $\Lambda$nDM, $w$CDM, IDE, and HDE, together with their open versions where $\Omega_K$ is released as a free parameter. For each model, we report the model selection indicators ($\Delta$AIC, $\Delta$DIC) with respect to the $\Lambda$CDM model.}
\label{tab:results}
\setlength{\tabcolsep}{2mm}
\renewcommand{\arraystretch}{1.4}
\begin{center}{\centerline{
\begin{tabular}{lllllcc}
\bottomrule[1pt]
Model   & Parameter     &       &                   &                                                         &$\Delta{\rm AIC}$       &$\Delta{\rm DIC}$  \\
\bottomrule[1pt]
{$\Lambda$CDM}
        &$H_0=70.0\pm2.2$                   &$\Omega_{\rm m}=0.310\pm0.008 $                        &                                                                       &                                                               &$0$                       &$0$            \\
{o$\Lambda$CDM}
        &$H_0=69.8\pm2.2$                   &$\Omega_{\rm m}=0.294\pm0.012$                         &                                                                       &$\Omega_K=0.049\pm0.037$                       &$0.577$            &$0.323$    \\
{$\Lambda$nDM}
        &$H_0=69.6\pm2.3$                    &$\Omega_{\rm m}=0.348^{+0.022}_{-0.026}$          &$w_{\rm dm}=-0.045\pm0.023$                            &                                                               &$-1.602$               &$-2.085$    \\
{o$\Lambda$nDM}
        &$H_0=69.6\pm2.3$                    &$\Omega_{\rm m}=0.345^{+0.034}_{-0.038}$          &$w_{\rm dm}=-0.041\pm0.028$                            &$\Omega_K=0.006\pm0.047$                       &$0.728$            &$-0.157$       \\
{$w$CDM}
        &$H_0=69.6\pm2.2$                    &$\Omega_{\rm m}=0.297\pm0.009$                        &$w_{\rm de}=-0.912^{+0.038}_{-0.035}$                &                                                                 &$-2.755$           &$-3.438$        \\
{o$w$CDM}
        &$H_0=69.6\pm2.3$                    &$\Omega_{\rm m}=0.298\pm0.012$                        &$w_{\rm de}=-0.914^{+0.049}_{-0.042}$                &$\Omega_K=-0.001\pm0.046$                    &$-1.705$           &$-0.519$          \\
{IDE}
        &$H_0=69.5\pm2.2$                    &$\Omega_{\rm m}=0.358\pm0.024$                        &$\beta=-0.26\pm0.11$                                           &                                                               &$-2.894$           &$-3.269$          \\
{oIDE}
        &$H_0=69.5\pm2.3$                    &$\Omega_{\rm m}=0.357\pm0.037$                        &$\beta=-0.25^{+0.12}_{-0.15}$                              &$\Omega_K=0.002\pm0.046$                       &$-0.885$           &$-1.602$          \\
{HDE}
        &$H_0=69.7\pm2.2$                    &$\Omega_{\rm m}=0.271\pm0.009$                        &$c=1.004^{+0.067}_{-0.081}$                                &                                                               &$-1.264$               &$-1.383$          \\
{oHDE}
        &$H_0=69.5\pm2.2$                    &$\Omega_{\rm m}=0.271\pm0.010$                        &$c=1.004\pm0.088$                                              &$\Omega_K=-0.004\pm0.045$                     &$0.509$          &$0.374$       \\
\bottomrule[1pt]
\end{tabular}}}
\end{center}
\end{table*}

Within the o$\Lambda$CDM model, individual probes favor spatial flatness: BAO, SN (DES-Dovekie), CC, and TD give $\Omega_K = 0.025\pm0.041$, $0.15\pm0.15$, $0.06\pm0.47$, and $-0.27^{+0.31}_{-0.39}$, respectively. Three supernova compilations are all consistent with flatness: DES-Dovekie yields $\Omega_K=0.15\pm0.15$, Pantheon+ yields $\Omega_K = 0.10\pm0.13$, and Union3 yields $\Omega_K=0.20\pm0.18$ (which lies slightly above $1\sigma$ but is statistically insignificant). We adopt DES-Dovekie (with the most recent photometric calibration) for our main analysis, and use Pantheon+ and Union3 for robustness checks.

However, the joint constraint favors an open Universe, as shown in Fig.~\ref{Contours}, consistent with the earlier finding of Ref.~\cite{Wu:2024faw} ($\Omega_K = 0.106 \pm 0.056$, a $1.9\sigma$ deviation from flatness). Our combined data yield $\Omega_K = 0.049 \pm 0.037$, a $1.3\sigma$ deviation from flatness, providing weaker support for an open Universe than previously claimed. This result differs from the Planck PR4 CMB result ($\Omega_K = -0.012 \pm 0.010$ \cite{Tristram:2023haj}) at $1.6\sigma$ tension. Replacing the supernova sample with Pantheon+ or Union3 gives $\Omega_K = 0.046 \pm 0.036$ and $\Omega_K = 0.046 \pm 0.039$, respectively, both still favoring an open geometry at comparable significance. That each individual probe is consistent with flatness while their combination favors an open Universe is not contradictory: the two most constraining cosmological probes (BAO and SN) both return positive $\Omega_K$ central values; when combined, the central value falls between the individual peaks while the error bars shrink, resulting in an open-Universe preference.

The above analysis is restricted to o$\Lambda$CDM. However, growing evidence suggests new physics beyond this paradigm. If $\Lambda$CDM is incomplete, curvature constraints derived within o$\Lambda$CDM could be systematically biased or even physically misleading. To assess the robustness of our measurements, we extend o$\Lambda$CDM by considering four classes of new physics, each introducing a single additional degree of freedom that exhibits a non-trivial degeneracy with $\Omega_K$. These extensions are: (i) non-cold dark matter, parameterized by $w_{\rm dm}$; (ii) dynamical dark energy, parameterized by $w_{\rm de}$; (iii) interacting dark energy, parameterized by $\beta$; (iv) holographic dark energy, parameterized by $c$.

Before measuring the curvature parameter in these extended frameworks, we first check whether the data prefer new physics assuming a flat universe. The constraint results are shown in Fig.~\ref{Contours}. Allowing non-zero dark matter EoS gives $w_{\rm dm}=-0.045\pm0.023$ ($2\sigma$ deviation from cold dark matter). Dynamical dark energy model yields $w_{\rm de}=-0.912^{+0.038}_{-0.035}$ ($2.5\sigma$ deviation from the cosmological constant). Interacting dark energy model gives $\beta=-0.26\pm0.11$ ($2.4\sigma$ evidence for dark energy decaying into dark matter). These results suggest that flat $\Lambda$CDM may be inadequate for late-time evolution. For HDE, we find $c=1.004^{+0.067}_{-0.081}$, consistent with the $\Lambda$CDM model (since HDE reduces to $\Lambda$CDM when $c=1$). However, recent studies indicate that HDE faces observational challenges and tensions between early- and late-Universe constraints \cite{Li:2024qus,Wu:2025vfs,Li:2024bwr}.

We constrain $\Omega_K$ within extended models; the results are presented in Fig.~\ref{Contours}. As can be seen, $w_{\rm dm}$ and $\beta$ are positively correlated with $\Omega_K$, while $w_{\rm de}$ and $c$ are negatively correlated. In all cases, the derived $\Omega_K$ values are consistent with spatial flatness: $0.006\pm0.047$ (o$\Lambda$nDM), $-0.001\pm0.046$ (o$w$CDM), $0.002\pm0.046$ (oIDE), and $-0.004\pm0.045$ (oHDE). Notably, the preference for new physics persists even when curvature is allowed: the additional parameters remain $w_{\rm dm}=-0.041\pm0.028$ for the o$\Lambda$nDM model, $w_{\rm de}=-0.914^{+0.049}_{-0.042}$ for the o$w$CDM model, and $\beta=-0.25^{+0.12}_{-0.15}$ for the oIDE model, respectively. Hence, the data do not genuinely favor an open Universe. The $\Omega_K>0$ preference in o$\Lambda$CDM arises from its limited flexibility: freeing curvature absorbs effects that physically belong to extended dynamics. Once extra parameters are introduced, the curvature signal disappears and the data instead favor flatness. Thus, the open-Universe hint is an artifact of over-restrictive assumptions, and curvature constraints from minimal o$\Lambda$CDM should be interpreted with caution.

\begin{figure}
\includegraphics[scale=0.33]{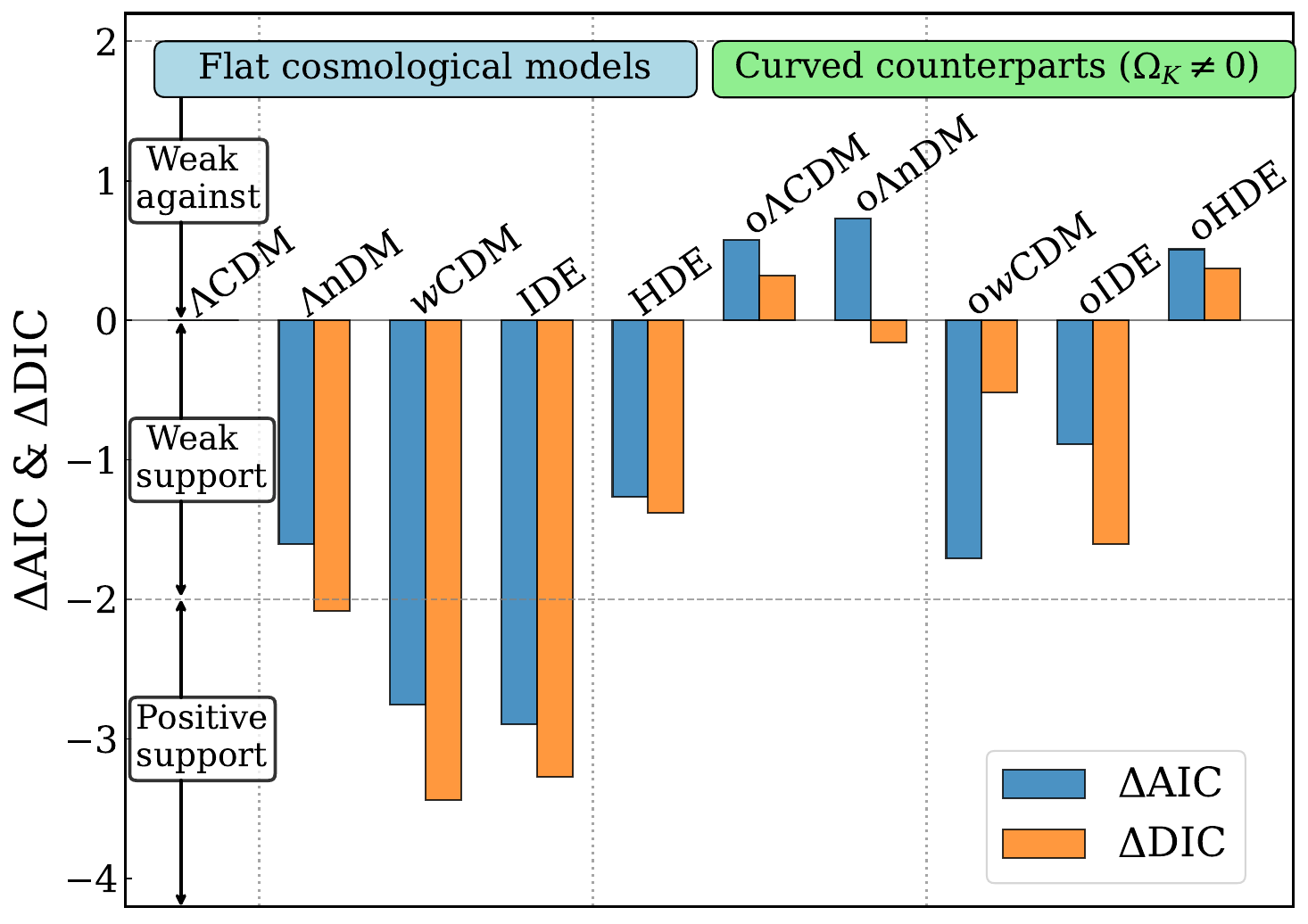}
\centering
\caption{Model comparison results for flat and non-flat $\Lambda$CDM and its extensions. The $\Delta$AIC and $\Delta$DIC values are relative to flat $\Lambda$CDM. Negative values indicate preference for the extended model; positive values favor flat $\Lambda$CDM.}
\label{AICDIC}
\end{figure}

To further verify whether the data favor new physics over an open Universe, we perform a model comparison using the Akaike Information Criterion (AIC) and Deviance Information Criterion (DIC). These are defined as
\begin{align}
\text{AIC} = \chi^2_{\text{min}} + 2k,\ \ \ \text{DIC} = \chi^2_{\text{min}} + 2k_{\text{eff}},
\end{align}
where $k$ is the number of free parameters and $k_{\text{eff}} = \langle \chi^2 \rangle - \chi^2_{\text{min}}$ is the effective number of constrained parameters. Lower AIC or DIC values indicate a better fit. We adopt the flat $\Lambda$CDM model as the baseline, and evaluate $\Delta\text{AIC} = \text{AIC}_{\text{model}} - \text{AIC}_{\Lambda\text{CDM}}$ (and analogously for $\Delta\text{DIC}$). Negative $\Delta$ values therefore indicate that the extended model is preferred over $\Lambda$CDM, while positive values disfavor it. Following the standard interpretation, $|\Delta| \in [0,2)$ suggests weak evidence, $[2,6)$ denotes positive evidence, and $>6$ means strong evidence, with the sign indicating which model is favored. The results of model comparison are presented in Fig.~\ref{AICDIC}.

As can be seen, all flat extensions are preferred over the $\Lambda$CDM model, with evidence ranging from weak to positive. In contrast, o$\Lambda$CDM is slightly disfavored relative to $\Lambda$CDM. Freeing the curvature parameter does reduce $\chi^2_{\rm min}$, but the improvement is too small to offset the penalty for the extra parameter in AIC and DIC. In contrast, the flat extensions achieve a sufficiently large reduction in $\chi^2_{\rm min}$ through additional physical effects, so they remain favored even after penalization. Moreover, for each extension, the flat version performs better than its non-flat counterpart. These results indicate that physically motivated extensions are more effective at improving the fit than simply allowing curvature. The preference for an open Universe under the minimal $\Lambda$CDM framework is therefore an artifact of its limited flexibility. Hence, the estimate $\Omega_K = 0.049 \pm 0.037$ obtained within $\Lambda$CDM is likely a framework-dependent result, and the data do not genuinely favor a curved Universe.

\section{Conclusion}
Cosmic curvature is deeply connected to the inflationary paradigm and the ultimate fate of the Universe. Using improved late-Universe observations, we find that the mild preference for an open Universe within the $\Lambda$CDM framework ($\Omega_K=0.049\pm0.037$, a $1.3\sigma$ hint) disappears once the model is extended to include new physics. In particular, $\Lambda$CDM extensions featuring non-cold dark matter, dynamical dark energy, or interaction between dark sectors all favor spatial flatness and provide $>2\sigma$ support for new physics. Model comparison further reveals that flat extensions are preferred over $\Lambda$CDM, and each flat extension outperforms its curved counterpart. Therefore, the open-Universe signal may not be a genuine feature of the observational data; rather, it is an artifact of the limited flexibility of the minimal $\Lambda$CDM model. Admittedly, our results, while supported by the new data, do not reach a definitive statistical significance. Future high-precision data will be essential to confirm or refute the interpretations presented here \cite{Wu:2022jkf}. It is worth noting that the CMB data, when interpreted within the $\Lambda$CDM framework, show a mild tendency toward a closed Universe ($\Omega_K<0$). However, this preference is itself derived under the $\Lambda$CDM framework, and may likewise be an artifact of its restrictive assumptions. If the standard cosmological paradigm is indeed incomplete, then any curvature constraints derived under its umbrella are inherently fragile. Future work using early-Universe and full-redshift data is necessary to constrain the curvature parameter under various $\Lambda$CDM extensions.

\begin{acknowledgments}
This work was supported by the Natural Science Foundation of Ningxia Province, China (No.~2026AAC030079) and the Research Starting Funds for Imported Talents, Ningxia University (Grant No. 030700002562).
\end{acknowledgments}

\section{Data Availability}
The data that support the findings of this article are openly available \cite{DESI:2025zgx,Brout:2022vxf,DES:2025sig,Rubin:2023jdq,Moresco:2022phi,Loubser:2025fzl,TDCOSMO:2025dmr}.

\bibliography{curvature}

\end{document}